\documentclass[reprint,showpacs,preprintnumbers,amsmath,amssymb,prc,floatfix]{revtex4-1}
\usepackage{float}
\usepackage{color}
\usepackage{graphicx}
\usepackage{dcolumn}
\usepackage{bm}
\usepackage{xcolor}
\usepackage{multirow}
\usepackage{comment}
\usepackage[colorlinks,citecolor=blue,linkcolor=red,anchorcolor=blue,filecolor=blue,urlcolor=blue]{hyperref}

\begin{document}
\title{Preformation Probability  and Kinematics of Clusters Emission yielding Pb-daughters}

\author{Joshua T. Majekodunmi$^1$}
\email{majekjoe1@gmail.com}
\author{M. Bhuyan$^{2,3}$}
\email{bunuphy@um.edu.my}
\author{K. Anwar$^1$}
\author{N. Abdullah$^1$}
\author{Raj Kumar$^4$}
\email{rajkumar@thapar.edu}
\affiliation{$^1$Institute of Engineering Mathematics, Universiti Malaysia Perlis, Arau, 02600, Perlis, Malaysia}
\affiliation{$^2$Center for Theoretical and Computational Physics, Department of Physics, Faculty of Science, University of Malaya, Kuala Lumpur 50603, Malaysia}
\affiliation{$^3$Institute of Research and Development, Duy Tan University, Da Nang 550000, Vietnam}
\affiliation{$^4$School of Physics and Materials Science, Thapar Institute of Engineering and Technology, Patiala, Punjab 147004, India}

\date{\today}

\begin{abstract}
\noindent
In the present study, the newly established preformation formula is applied for the first time to study the kinematics of the cluster emission from various radioactive nuclei, especially those decaying to the double-shell closure $^{208}$Pb nucleus and its neighbours as daughters. The recently proposed universal cluster preformation formula has been established based on the concepts that underscore the influence of the mass and charge asymmetry ($\eta_A$ and $\eta_Z$), cluster mass $A_c$ and the Q-value, paving the way to quantify the energy contribution during the preformation as well as the tunnelling process separately. The cluster-daughter interaction potential is obtained by folding the relativistic mean-field (RMF) densities with the recently developed microscopic R3Y using the NL$3^*$ and the phenomenological M3Y NN potentials to compare their adaptability. The penetration probabilities are calculated from the WKB approximation. With the inclusion of the new preformation probability $P_0$, the predicted half-lives from the R3Y and M3Y interactions are in good agreement with the experimental data. Furthermore, a careful inspection reflects slight differences in the decay half-lives, which arise from their respective barrier properties. The $P_0$ for the systems with the double magic shell closure $^{208}$Pb daughter are found to be relatively higher with an order of $\approx 10^2$ than those with neighbouring Pb-daughter nuclei. By exploring the contributions of the decay energy, the recoil effect of the daughter nucleus is appraised, unlike several other conjectures. Thus, the centrality of the Q-value in the decay process is demonstrated and re-defined within the preformed cluster-decay model. Besides, we have introduced a simple and intuitive set of criteria that governs the estimation of recoil energy in the cluster radioactivity. \\
\par 
\end{abstract}
\maketitle

\section{INTRODUCTION} \label{sec:level1}
\noindent
Clustering is one of the notable dynamical attributes of an atomic nucleus exhibiting regular patterns, despite the complexities associated with the nuclear many-body systems. Its prediction dates back to the theoretical investigation of  Sandulescu \textit{et al.} in 1980 \cite{sand80} in which the shell closure effect of one of the reactants was used to reproduce a cold reaction based on the fragmentation theory \cite{gupt94}. The findings were subsequently validated by Rose and Jones \cite{rose84} where cluster radioactivity was established as a highly asymmetric spontaneous disintegration of radioactive nuclei in which the emitted particle is heavier than $^4$He and yet smaller than the lightest fission fragments. So far, the emission of  $^{14}$C, $^{18,20}$O, $^{23}$F, $^{22,24-26}$Ne, $^{28,30}$Mg and $^{32,34}$Si clusters from various trans-lead nuclei ($^{221}$Fr - $^{242}$Cm) have been observed \cite{bone07}. The emitted clusters in this region are usually associated with the double magic nucleus $^{208}_{82}$Pb or nuclei in its vicinity as daughters. Considering the kinematics of cluster emission, earlier studies \cite{qian12,deng15} have revealed that like $\alpha$-decay and spontaneous emission,  the rate at which clusters are emitted from odd-parent nuclei is confronted with more structural hindrance as compared to the rate of cluster emissions from its neighbouring even-even isotopes. Moreover, it has been shown that there is a substantial difference between the observed kinetic energy and Q-value of the cluster decay, suggesting a considerable recoil effect \cite{hoos05}.

From the theoretical viewpoint, cluster radioactivity follows the description of the Gamow model of $\alpha$-decay which hinges on the quantum tunnelling effect. This description can be grouped into two main categories based on their treatment of cluster emission namely fission and $\alpha$-like models. The fission models e.g the analytic super asymmetric fission model (ASAFM) of Poenaru {\it et al.} \cite{poen85,poen86} assume that the parent nucleus undergoes continuous deformation until it penetrates the confining interaction barrier and thus attains the saddle configuration. This approach takes no cognizance of the preformation of the cluster within the parent nucleus before its emission. In other words, preformation probability ($P_0$) is taken as unity.  In contrast, the $\alpha$-like models like the preformed cluster-decay model (PCM) \cite{mali89,gupt88,wei17}, which is rooted in the quantum mechanical fragmentation theory (QMFT), assume that clusters are composed of several nucleons pre-born within the parent nucleus before tunnelling through the potential barrier. Thus, realistic values of $P_0$ can be calculated and as a result, the experimental half-lives can be accurately reproduced. The literature \cite{blen88,sant21c,ni10,bala14,deng14} is replete with different expressions to estimate $P_0$ of which the predictive power of most are either restricted to a certain region of the nuclear chart or fitted with some arbitrary constants with no clear link/relevance to the kinematics and underlying concept of cluster emission.

Deliberate attention has been given to this concept in the derivation of our newly proposed preformation formula \cite{josh22L} and is extended in the present study to reproduce the experimentally measured cluster decay half-lives. Besides, by exploring the mechanism and kinematics in the decay channel, we contemplate that a certain amount of energy must be expended during cluster formation just before its emission, unlike previous studies \cite{ropk14,xu16} where much emphasis was given to the kinetic energy but the recoil energy was assumed to be negligible. Assumption such as Levinger’s approximation \cite{levi53} where the recoil energy is considered too small becomes invalid for natural radioactivity in which heavy ions are emitted \cite{stra01}. Hence, this study is aimed at investigating the systematic contribution of the decay energy into three distinct parts, accounting for cluster preformation, emission or tunnelling and the residual energy with which the daughter nucleus recoils. Also, the relative separation between the decay fragments denoted $\Delta$R is used to account for the neck-formation effect which decides the first turning point for the barrier penetration within the PCM \cite{kuma12c} which is employed here for the present investigation. This barrier is formed by the interplay of the Coulomb and nuclear potential.

The Coulomb potential can be simply estimated as the ratio of the product of the charges of the decay fragments (emitted cluster and the daughter nucleus) to the sum of their radii. However, obtaining the nuclear potential usually involves the use of either the phenomenological \cite{quen78,horn75} or microscopic approaches \cite{schu16,vaut72}.  Besides the fundamental approaches \cite{epel09,ekst13}, the R3Y nucleon-nucleon (NN) potential \cite{sing12,sing10} which stems from the relativistic mean-field (RMF) Lagrangian using the NL$3^*$ parameter set is employed in the present study along with the phenomenological M3Y NN potential \cite{satc79}. The RMF theory is apt to take care of the ground and excited-state properties of the atomic nuclei \cite{Bisw20,Itag20,Tani20}. The Q-values are also calculated from RMF (NL$3^*$) and are compared with the macroscopic-microscopic WS3 \cite{liu11} and those obtained from the experimental binding energy data \cite{wang17}. The WKB approximation is used to estimate the penetration probability $P$.  The paper is presented in the following manner: Section \Ref{theory} gives a brief description of the theoretical framework which includes the relativistic mean-field (RMF) and the folding procedure for R3Y and M3Y NN potentials. The preformed cluster-decay model (PCM) and the new $P_0$ formula yielding a set of new equations are also presented. Section \Ref{result} details the presentation of the results and their corresponding discussions. Finally, the conclusion and summary of this work are given in Section \Ref{summary}.

\section{Theoretical formalism}
\label{theory} \noindent
The relativistic mean-field (RMF) Lagrangian is built from the interaction between the nucleonic field and the three mesonic fields, isoscalar–scalar $\sigma$, isoscalar–vector $\omega$, and isovector–vector $\rho$ as well as the photon field $A_\mu$ together with their respective coupling constants ($g_\sigma$, $g_\omega$, $g_\rho$), is given by \cite{jos22a,ring96,bhu15,bhu20}
\begin{eqnarray}
{\cal L}&=&\overline\psi_i\left\{i\gamma^{\mu}\partial_{\mu}-M\right\}\psi_i+\frac{1}{2}\partial^{\mu}\sigma\partial_{\mu}\sigma\nonumber\\
&&-\frac{1}{2}m_{\sigma}^{2}\sigma^2-\frac{1}{3}g_2\sigma^3-\frac{1}{4}g_3\sigma^4-g_\sigma\overline{\psi}_i\psi_i\sigma\nonumber\\
&&-\frac{1}{4}\Omega^{\mu\nu}\Omega_{\mu\nu} +\frac{1}{2}m^2_\omega \omega^\mu \omega_\mu-g_\omega\overline{\psi}_i\gamma^\mu\psi_i \omega_\mu\nonumber\\
&&-\frac{1}{4}\vec B^{\mu\nu}.\vec B_{\mu\nu}+\frac{1}{2}m^2_\rho\vec \rho^\mu.\vec \rho_\mu-g_\rho\overline{\psi}_i\gamma^\mu\vec{\tau}\psi_i.\vec \rho^\mu\nonumber\\
&&-\frac{1}{4}F^{\mu\nu}F_{\mu\nu}-e\overline{\psi}_i\gamma^\mu(\frac{1-\tau_{3i}}{2})\psi_i A_\mu. 
\label{lag}
\end{eqnarray}
Parameters $g_2$, $g_3$ and $\frac{e^2}{4\pi}$ are the coupling constants of the non-linear terms. The third component of the isospin is $\tau_{3i}$. $M$ is the mass of nucleons while the  masses of $\sigma$, $\omega$ and $\rho$-mesons are $m_\sigma$, $m_\omega$, $m_\rho$ and respective fields $\omega^{\mu}$, $\vec \rho_\mu$ and $A_\mu$. It is worth noting that the contribution of the $\pi$-meson has been omitted in Eq. (\Ref{lag}) in the mean-field calculation as a result of its pseudoscalar nature \cite{ring96,sero86}. A detailed description of the field tensors for $\omega^{\mu}$, $\vec \rho_\mu$ and $A_\mu$ fields can be found in Ref \cite{sing22} and the references therein. The field tensors are treated as classical fields and thus, the Dirac equation is obtained for the nucleons and simplified as
\begin{equation}
    [-i\alpha.\nabla+\beta(M^*+g_\sigma\sigma)+g_\omega\omega+g_\rho\tau_3\rho_3]\psi_i=\epsilon_i\psi_i.\\
\end{equation}
Similarly, the Klein-Gordon equations for the participating mesons are simplified as
\begin{eqnarray}
 (-\nabla^2+m^2_\sigma)\sigma(r)&=&-g_\sigma\rho_s(r)-g_2\sigma^2(r)-g_3\sigma^3(r),\nonumber\\
   (-\nabla^2+m^2_\omega)V(r)&=&g_\omega\rho(r),\nonumber\\
   (-\nabla^2+m^2_\rho)\rho(r)&=&g_\rho\rho_3(r).
\end{eqnarray}
This equations are solved self consistently using the NL3$^*$ parameter set. Within the limit of one-meson exchange for a heavy and static baryonic medium, the microscopic R3Y NN potential is obtained as
\begin{eqnarray}
V_{eff}^{R3Y}(r)&=&\frac{g^2_\omega}{4\pi}\frac{e^{-m_\omega r}}{r}+\frac{g^2_\rho}{4\pi}\frac{e^{-m_\rho r}}{r}-\frac{g^2_\sigma}{4\pi}\frac{e^{-m_\sigma r}}{r}\nonumber\\
&&+\frac{g^2_2}{4\pi}re^{-2m_\sigma r} +\frac{g^2_3}{4\pi}\frac{e^{-3m_\sigma r}}{r} +J_{00}(E)\delta(s),
\label{r3y} 
\end{eqnarray}
where $J_{00}(E)\delta(s)$ is the zero-range pseudopotential denoting the exchange effect. Eq. (\Ref{r3y}) is similar to the phenomenological prescription of Reid-Elliott \cite{satc79} called   M3Y NN potential which is constructed to reproduce the G-matrix element. The  M3Y NN potential takes the form
\begin{equation}
V_{eff}^{M3Y}(r)=7999\frac{e^{-4r}}{4r}-2134\frac{e^{-2.5r}}{2.5r}+J_{00}(E)\delta(s).
\label{m3y}
\end{equation}
The double folding technique \cite{satc79} is employed to estimate the nuclear interaction potential $V_n(R)$ and expressed as
\begin{equation}
    V_n(R)=\int dr_c\int dr_d \rho_c(\vec r_c)\rho_d(\vec r_d)V_{eff}(\vec r_{cd}=\vec R+\vec r_d- \vec r_c),
    \label{fold}
\end{equation}
where $\rho_c$ and $\rho_d$ are the nuclear matter densities of the cluster and daughter nuclei. $V_n(R)$ given by Eq. (\Ref{fold}) combines with the Coulomb potential $V_C(R)=\frac{Z_{c}Z_d}{R}e^2$ to obtain the total interaction potential
\begin{equation}
    V(R)= V_n(R)+V_C(R),
\end{equation}
which is used to estimate the WKB penetration probability (as illustrated in Fig. \Ref{fig 1}) and hence, the cluster decay half-lives using the preformed cluster-decay model (PCM) \cite{kuma12c}. The penetration probability of clusters across the tunnelling path is  given as
\begin{equation}
    P=P_aW_iP_b
\end{equation}
which involves a three step process, shown and discussed in Fig. \Ref{fig 2}.
\begin{eqnarray}
    P_a&=&\exp\left(-\frac{2}{\hbar}\int^{R_i}_{R_a}\{2\mu[V(R)-V(R_i)]\}^{1/2}dR\right),\\
    \nonumber \mbox{and}\\
    P_b&=&\exp\left(-\frac{2}{\hbar}\int_{R_i}^{R_b}\{2\mu[V(R_i)-Q]\}^{1/2}dR\right).\label{wkb} 
\end{eqnarray}

\subsection{Preformed cluster-decay model (PCM)}
The decay half-life $T_{1/2}$ within the preformed cluster-decay model (PCM) can be defined in term of the decay constant $\lambda$, penetration probability $P$, and preformation probability $P_0$ with the expression
\begin{equation}
    T_{1/2} =\frac{\ln2}{\lambda}, \hspace{0.5cm} \lambda= \nu_{0} P_0 P. \label{11}
\end{equation}
The assault frequency $\nu_0$ has nearly constant value of $10^{21}$ s$^{-1}$ and can be calculated as 
\begin{equation}
    \nu_{0}=\frac{\mbox{ velocity }}{R_0}=\frac{\sqrt{2E_{c}/\mu}}{R_0}, \label{afreq}
\end{equation}
where $R_0$ denotes the radius of the parent nucleus and $E_{c}$ is the kinetic energy of the emitted cluster.  The Q-values are estimated from the ground state binding energies from RMF,  AME2016  \cite{wang17}, WS3 \cite{liu11} mass tables  using the expression
 \begin{equation}
     Q=BE_p-(BE_d+BE_c), \label{qval}
 \end{equation}
where $BE_p$, $BE_d$ and $BE_c$ are the binding energies of the parent, daughter nuclei and the emitted cluster respectively.

Instead of the primitive cluster-mass dependent preformation formula of Blendowske and Walliser \cite{blen88}, here, we have given a close attention to study the relationships among various theoretically established properties/factors that influences cluster preformation such as the cluster mass $A_c$ \cite{sing11}, mass and charge asymmetries $\eta_A=(A_d-A_c)/(A_d+A_c)$ and $\eta_Z=(Z_d-Z_c)/(Z_d+Z_c)$ (since the emission of the same cluster from different parent nuclei as well as different clusters from the same parent nucleus is an experimentally observed fact \cite{bone07,gupt94,bone99}), the relative separation between the centers of the  fragments $r_B=1.2(A_c^{1/3}+A_d^{1/3})$ \cite{deli09,qian12} and the Q-value \cite{isma14}. Hence, we have proposed a new $P_0$ formula \cite{josh22L},
\begin{equation}
     \log P_0 = -\frac{aA_c\eta_A}{r_B}-Z_c\eta_Z+bQ+c, \label{P0}
 \end{equation}\\
where $a$, $b$ and $c$ are the fitting parameters in Ref.\cite{josh22L}. 
The measure of accuracy is evaluated using the $\chi^2$  expression 
\begin{equation*}
    \chi^2=\sum^{n}_{i=1} \frac{\left[\log_{10}^{Expt.}T_{1/2}-\log_{10}^{cal}T_{1/2}\right]^2}{\log_{10}^{cal}T_{1/2}}
\end{equation*}
for 14 even-even nuclei and 5 odd-A nuclei whose values are also given in Table 1 of  Ref. \cite{josh22L}.As explained in the $3^{rd}$ footnote of the table, it is worth mentioning that only the experimentally measured systems yielding Pb-daughters are preferentially considered in this fundamental study. Besides, it is interesting to note that the third term on the right-hand side of Eq. (\Ref{P0}) opens a new window to probe the contributions of the decay energy. In other words, the term $bQ$ gives a quantitative description of the \textit{energy contributed in cluster formation}. Thus, for the first time, the Q-value is presented in terms of its usage/disbursement in the kinematics of cluster emission as
\begin{eqnarray}
Q=\overbrace{\underbrace{ bQ}_{\substack{\text{energy  }\\ \text{contributed in}\\ \text{cluster formation}}}+\underbrace{\kappa\sqrt{Q}}_{\substack{\text{energy }\\ \text{contributed in}\\ \text{cluster emission}}}}^{E_c}+\underbrace{E_d}_{\substack{\text {recoil}\\ \text{energy of}\\ \text{daughter nucleus}}} \label{qexpd}
\end{eqnarray}
where the $\kappa\sqrt{Q}$ is the \textit{energy contributed in cluster emission}. Further, following the work of Gupta \textit{et al.} \cite{sing11}, the kinetic energy of the emitted cluster is expressed as
\begin{eqnarray}
 E_c=\frac{A_d}{A}Q=bQ+\kappa\sqrt{Q}.
 \label{ke}
\end{eqnarray}
On little simplification, we get
\begin{equation}
    \kappa=\sqrt{Q}\left(\frac{A_d}{A}-b\right). \label{kap}
\end{equation}
The quantity $\kappa$ in Eq.(\ref{kap}) refers to the tunneling factor. Detailed explanation and implication of the newly derived Eq.s (\Ref{P0})-(\Ref{kap}) are typified, analysed and discussed in the subsequent section.

\section{CALCULATIONS AND DISCUSSIONS}
\label{result}
\begin{figure}
\includegraphics[scale=0.42]{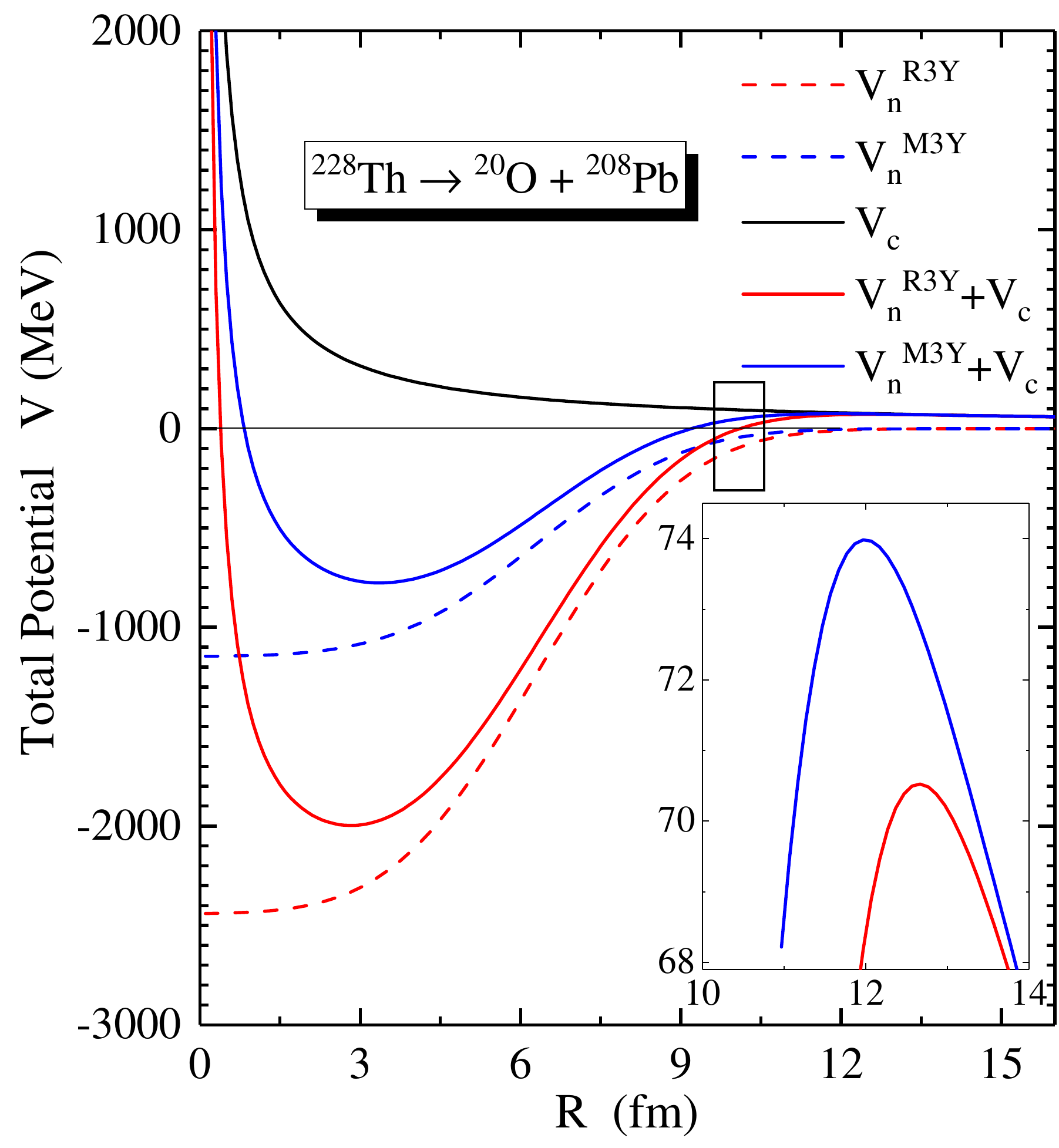} 
\caption{\label{fig 1} A schematic representation of the total nucleus-nucleus interaction potential V (MeV) and the respective contributions of the Coulomb and double-folded R3Y (NL$3^*$) and M3Y nuclear potentials as a function of the radial separation $R$ (fm) for a representative case of $^{228}$Th $\rightarrow ^{20}$O + $ ^{208}$Pb. The inset shows a magnified view of the barrier height and position.}
\end{figure}

The decay properties of $^{14}$C, $^{18,20}$O, $^{23}$F, $^{22-26}$Ne, $^{28-30}$Mg and $^{34}$Si clusters emitted from various heavy nuclei leading to the formation of daughters of Pb-isotopes. The effect of double-shell closure in terms of $^{208}$Pb daughter is analysed using the theoretical formalism discussed in the previous section. The relativistic mean-field theory (RMF) is employed here, being an efficient tool to reproduce the ground state properties of the decaying parent nuclei. The RMF-based R3Y (NL$3^*$) and the phenomenological M3Y NN interactions are folded with their respective RMF densities to deduce the nuclear interaction potential. As a representative case, Fig. \ref{fig 1} illustrates the individual contributions of the nuclear and Coulomb potentials which collectively forms the total interaction potential $V$ $(=V_c+V_n)$ for $^{228}$Th $\rightarrow ^{20}$O + $ ^{208}$Pb for the cases of R3Y (NL$3^*$) and M3Y interactions.
\begin{figure}
\includegraphics[scale=0.4]{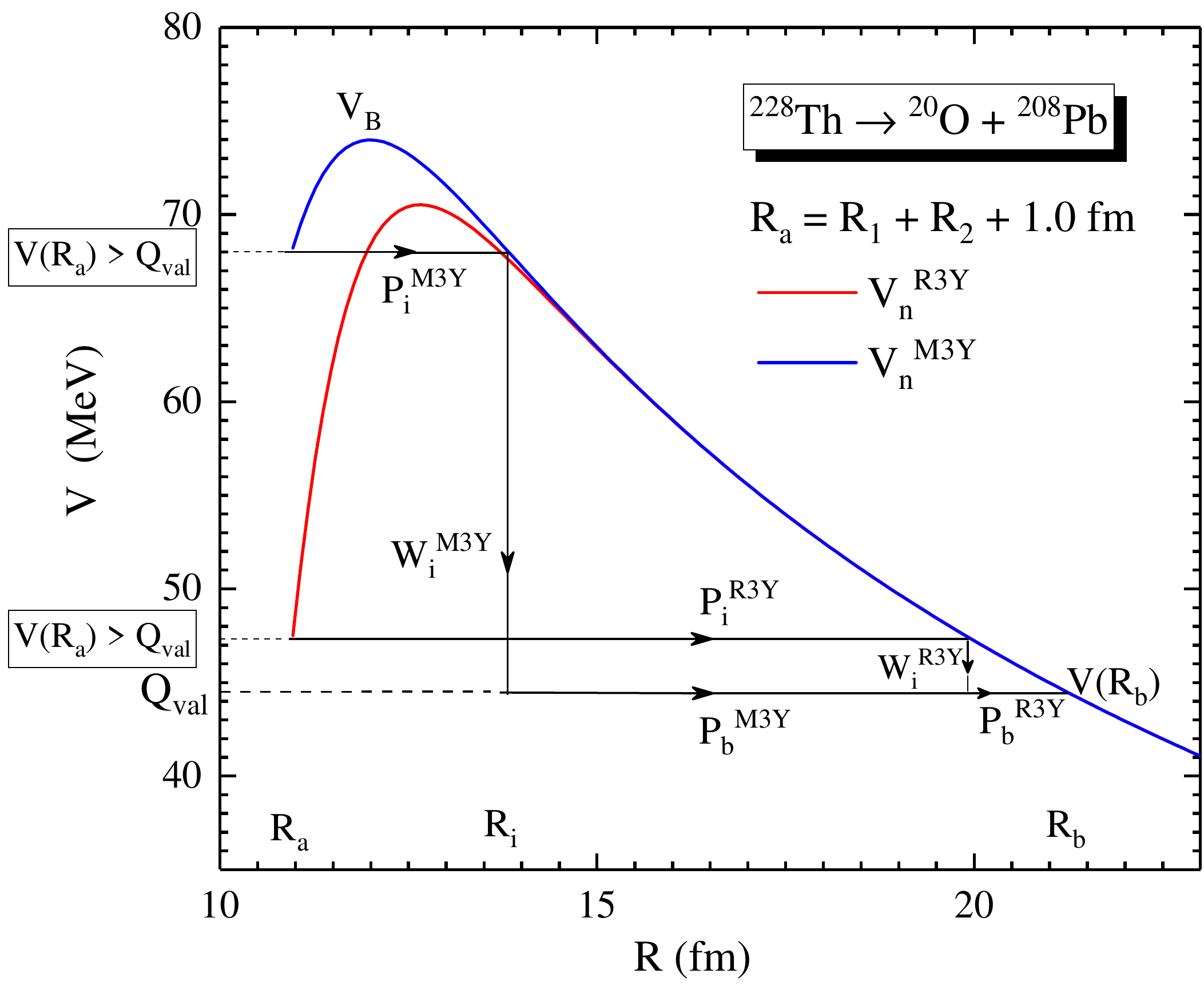} 
\caption{\label{fig 2} Total interaction potential for energetically favoured reaction $^{228}$Th $\rightarrow ^{20}$O + $ ^{208}$Pb for both M3Y and R3Y NN potentials, satisfying the criterion for cluster penetration $V(R_a)>Q_{val}$ at $\Delta$R = 1.0 fm (for relative comparison). The three step penetration process of interaction potential is also shown.}
\end{figure}

\subsection{Barrier Characteristics from M3Y and R3Y NN-Potentials}
From Fig. \ref{fig 1}, it is observed that the total interaction potential $V$ and nuclear potential $V_n (R)$ displays similar characteristics for both R3Y and M3Y NN potentials. However, the difference lies in their qualitative description which is apparent in the central region \cite{bhuy18,jos22a} but decreases proportionately with the radial separation $R$. Moreover, the R3Y and M3Y NN interactions are characterized by different barrier properties and hence the barrier height of the R3Y could be relatively lower (at about 3.6 MeV), being more attractive as shown in the inset.

The cluster penetration process for the energetically favoured reactions, taking $^{228}$Th $\rightarrow ^{20}$O + $ ^{208}$Pb as a representative case is shown in Fig. \ref{fig 2}. The three-step procedure involved is initiated by barrier penetration at the first turning point $R_a$ up to the point $R_i$ and followed by a de-excitation (given in the excitation model of Greiner and Scheid \cite{grei86}  for heavy  cluster emissions as $W_i = 1$) from $V(R_i)$ and thereafter, the cluster penetrates from $R_i$ to point $R_b$ such that $V(R_b)=Q_{val}$. It is imperative to note that this process is highly influenced by the decay energy  Q-value which must be positive. Although this description is not altogether new, the contribution of the Q-value to the energy contributed during cluster preformation (captured by the proposed preformation formula) is separately estimated/analysed for the first time. The figure further stresses the disparity in the barrier properties of the M3Y and R3Y NN potential as mentioned earlier. Here, for relative comparison, the neck-length parameter is kept at $\Delta$R = 1.0 fm for both NN potentials. Consequently, there is a significant difference between their respective barrier peak/height $V_B$ and its derivatives such as the barrier lowering parameter $\Delta V_B$ and the driving potential $V(R_a)-Q_{val}$.
\begin{figure}
\includegraphics[scale=0.4]{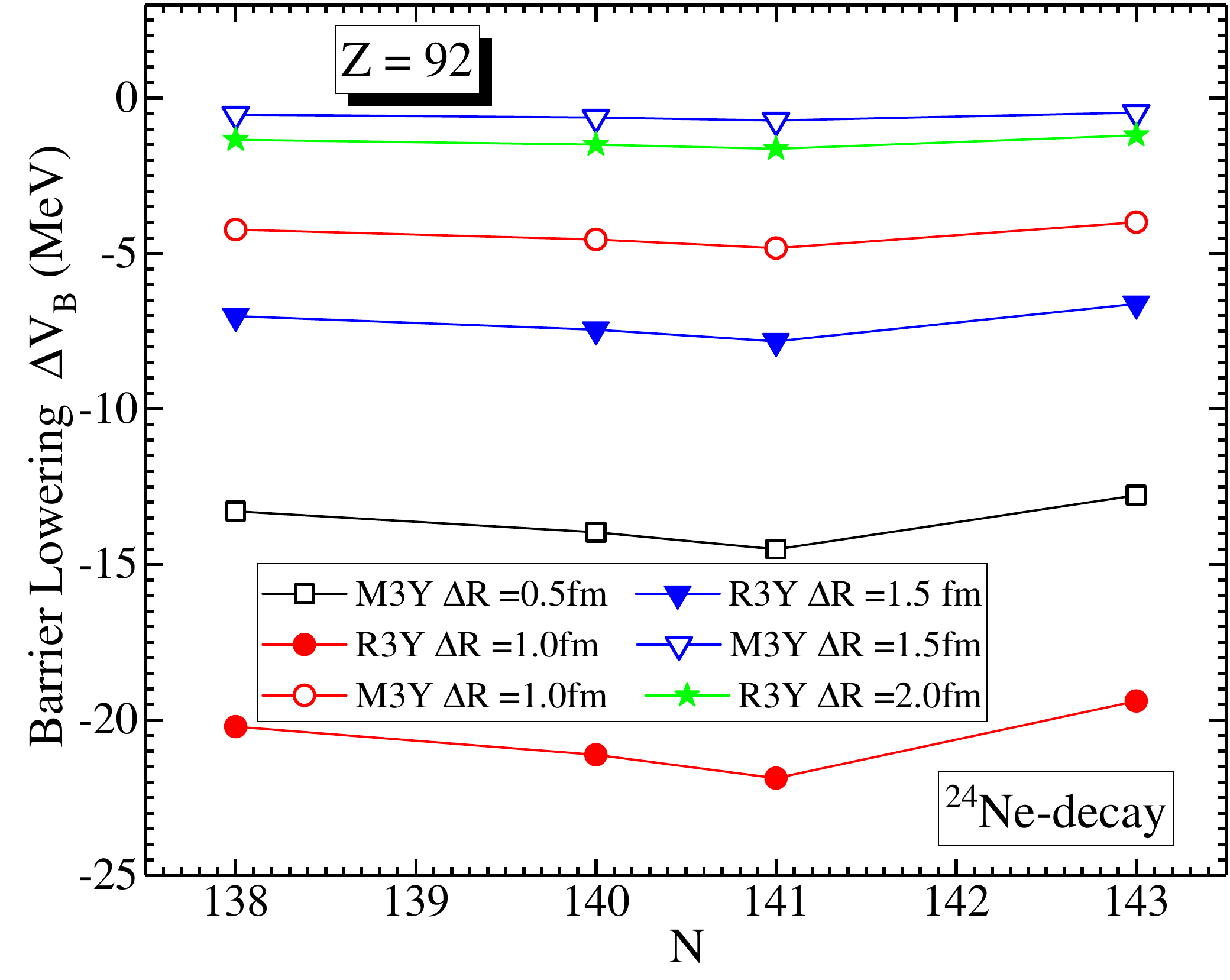}
\caption{\label{fig 3}The barrier lowering parameter $\Delta V_B$ (MeV) of $^{24}$Ne emission of $^{230,232,233,235}$U isotopes at different neck-length parameter $\Delta$R values.}
\end{figure}

The barrier lowering parameter $\Delta V_B$ is an inherent feature of the preformed cluster-decay model (PCM) which encapsulates various modifications in the barrier region especially those occasioned by the neck-length values \cite{shar19}. Fig. \ref{fig 3} illustrates  the profile of $\Delta V_B$ with respect to the neutron number N of $^{230,232,233,235}$U isotopes at varied neck-length parameter $\Delta$R values lying within the proximity potential limit \cite{bloc77,niyt15,shar18}. Considering the predictions from M3Y (open symbols) and R3Y (solid symbols) potentials separately, it is clear that the barrier lowering parameter is largely influenced by the neck configuration, and hence its modification increases with increasing $\Delta$R. Thus, $\Delta V_B$ dictates the cluster tunnelling path and it is usually negative since the penetration point is always below $V_B$.
\begin{figure}
\includegraphics[scale=0.4]{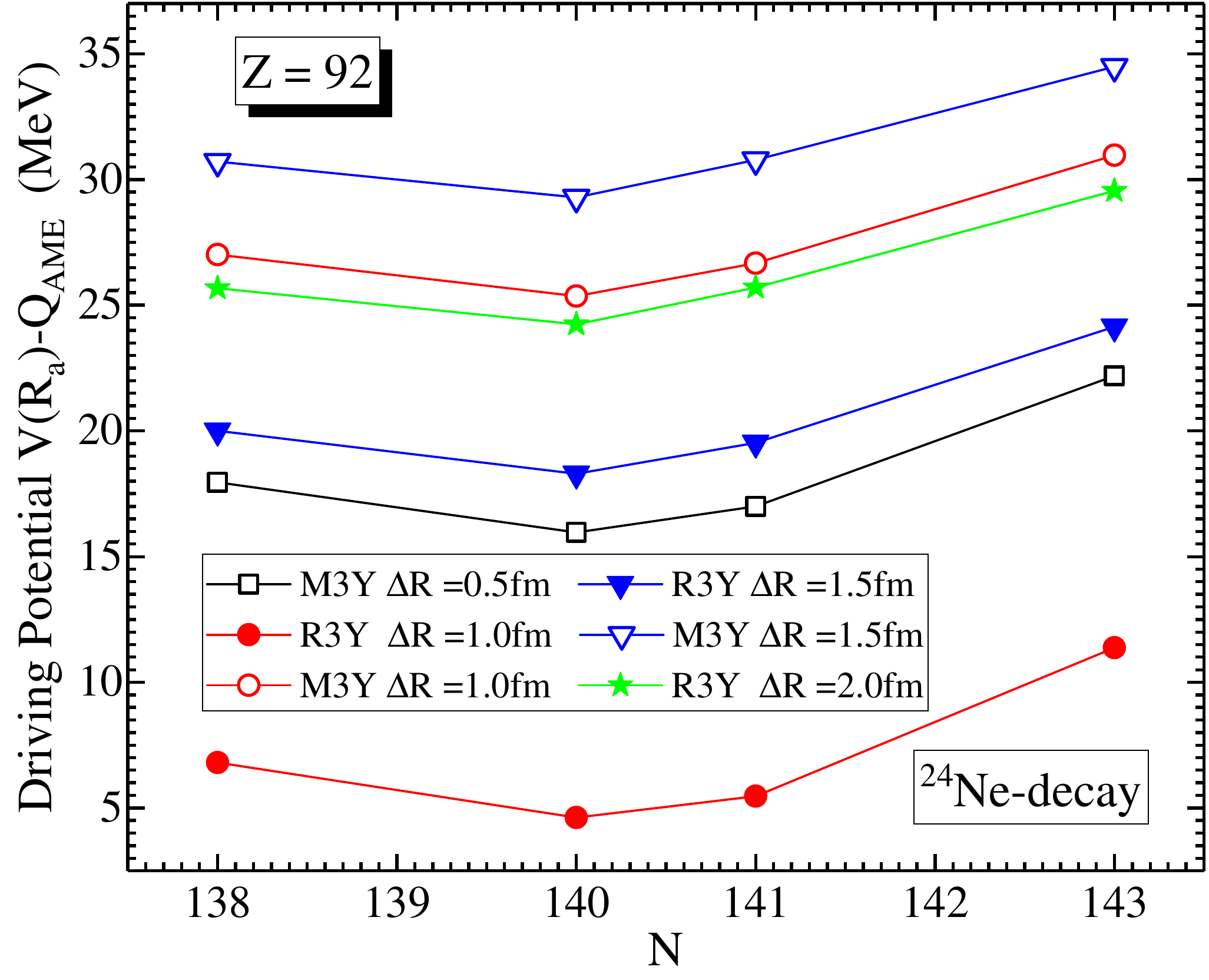}
\caption{\label{fig 4} The driving potential  $V(R_a)-Q_{AME}$ (MeV) of $^{24}$Ne emission as a function of the neutron number of $^{230,232,233,235}$U isotopes at different neck-lengths ($\Delta$R).}
\end{figure}

Similarly, the difference between the interaction potential and the energy available for the cluster decay process is referred to as the driving potential $V_d$ $(=V(R_a)-Q_{AME}$). It is worth mentioning that the Q-value ($Q_{AME}$) used here are calculated from the experimental binding energy data \cite{wang17} only for the sake of accuracy. In several studies, the minima in $V_d$ are usually used to indicate the most probable decay channels, here, variations are made in order to decide on the most appropriate neck length, especially for the recently developed R3Y interaction.  We have earlier demonstrated that $\Delta$R = 0.5 fm \cite{kuma12c,jos21c} is suitable enough for M3Y interaction in cluster decay studies. Nonetheless, this $\Delta$R value is not energetically favourable in the case of R3Y due to its unique barrier characteristics (see Ref.\cite{jos22b} for elaborate details on the range of predictability of both R3Y and M3Y interactions). Despite the change in $\Delta$R, a regular pattern is maintained in the profile of the driving potential for both interactions. Other than these variations, $\Delta$R is fixed at 1.0 fm for R3Y in the remaining part of this paper for the sake of relative comparison.
\begin{figure}
\includegraphics[scale=0.238]{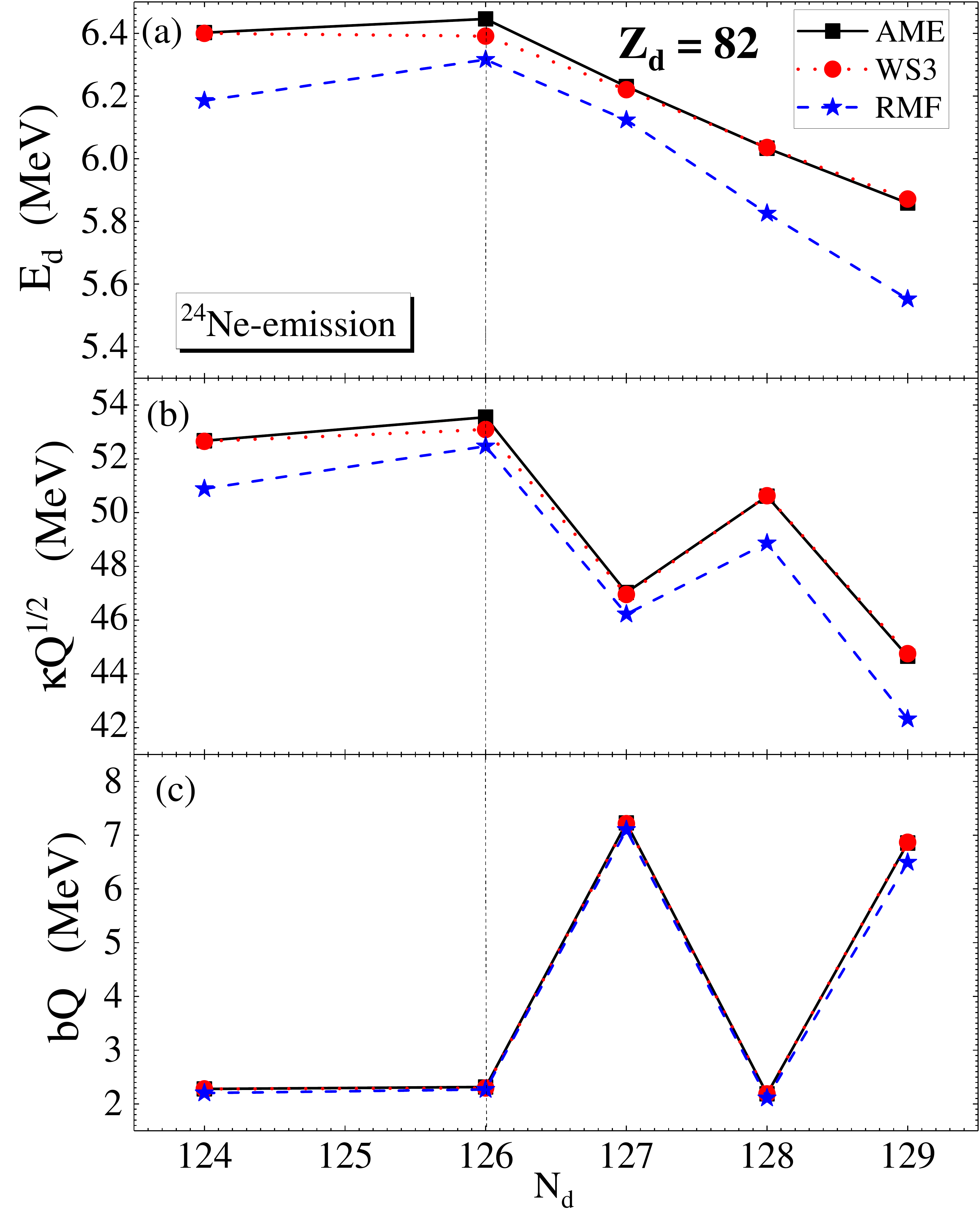}
\caption{\label{fig 5} Variation of the preformation properties: (a)  recoil energy of the daughter nuclei (b) cluster emission energy and (c) weighted Q-values of various Radium and Uranium isotopes as a function of the neutron number of the daughters formed from $^{A}U\rightarrow ^{24}Ne+^{A-24}Pb$.}
\end{figure}

\begin{table*}
\caption{\label{tab2} The decay half-lives and preformation analysis of various experimentally observed clusters \cite{bone07} forming different Pb-daughters. The Q-values are calculated using the binding energies from experimental data $(Q_{AME})$\cite{wang17}. Columns (8-13) are exclusively devoted to the cluster preformation and emission details from the proposed Eqs. (\Ref{P0}) - (\Ref{kap}).}
\begin{ruledtabular}
\begin{tabular}{ccccccccccccc}
Parent	&	Cluster	&	Daughters&	$Q_{AME}$&\multicolumn{3}{c}{$\log_{10}T_{1/2}$}	&$P_0$&	bQ&	$\kappa$	&$\kappa\sqrt{Q}$	& $E_{c}$ &	$E_{d}$\\
\cline{5-7}
nuclei&cluster&nuclei&(MeV) &Expt.&M3Y&R3Y& Eq. (\Ref{P0})&(MeV)& &(MeV)&(MeV)&(MeV)\\
\hline
$^{221}$Ra	&$^{14}$C	&$^{207}$Pb	&32.40	&13.39	&13.95	&14.22	&$9.19\times10^{-18}$&	3.86	&4.65	&26.49	&30.34	&2.05\\
$^{222}$Ra	&$^{14}$C	&$^{208}$Pb	&33.05	&11.01	&11.12	&11.94	&$8.82\times10^{-16}$&	1.22	&5.17	&29.74	&30.96	&2.08\\
$^{223}$Ra	&$^{14}$C	&$^{209}$Pb	&31.83	&15.06	&14.71	&15.40	&$8.22\times10^{-18}$&	3.79	&4.62	&26.04	&29.83	&2.00\\
$^{224}$Ra	&$^{14}$C	&$^{210}$Pb	&30.53	&15.86	&16.58	&17.10	&$6.71\times10^{-18}$&	1.13	&4.98	&27.50	&28.63	&1.91\\
$^{226}$Ra	&$^{14}$C	&$^{212}$Pb	&28.20	&21.19	&20.31	&20.60	&$5.68\times10^{-18}$&	1.04	&4.78	&25.41	&26.45	&1.75\\
$^{226}$Th	&$^{18}$O	&$^{208}$Pb	&45.73	&$\textgreater15.30$	&17.31	&17.63	&$9.84\times10^{-20}$&	1.69	&5.97	&40.39	&42.09	&3.64\\
$^{228}$Th	&$^{20}$O	&$^{208}$Pb	&44.72	&20.72	&20.54	&22.28	&$3.78\times10^{-21}$&	1.65	&5.85	&39.15	&40.80	&3.92\\
$^{230}$U	&$^{22}$Ne	&$^{208}$Pb	&61.39	&19.57	&19.91	&19.48	&$4.46\times10^{-23}$&	2.27	&6.80	&53.25	&55.52	&5.87\\
$^{231}$Pa	&$^{23}$F	&$^{208}$Pb	&51.89	&26.02	&25.67	&25.34	&$5.65\times10^{-26}$&	6.17	&5.63	&40.55	&46.72	&5.17\\
$^{230}$U	&$^{24}$Ne	&$^{206}$Pb	&61.35	&$\textgreater18.20$	&23.68	&23.18	&$2.47\times10^{-26}$&	2.27	&6.73	&52.68	&54.95	&6.40\\
$^{232}$U	&$^{24}$Ne	&$^{208}$Pb	&62.31	&21.08	&20.58	&20.94	&$2.92\times10^{-24}$&	2.31	&6.79	&53.56	&55.86	&6.45\\
$^{233}$U	&$^{24}$Ne	&$^{209}$Pb	&60.49	&24.84	&24.39	&24.55	&$2.30\times10^{-26}$&	7.20	&6.05	&47.06	&54.26	&6.23\\
$^{234}$U	&$^{24}$Ne	&$^{210}$Pb	&58.83	&25.92	&26.11	&26.11	&$1.97\times10^{-26}$&	2.18	&6.60	&50.62	&52.79	&6.03\\
$^{235}$U	&$^{24}$Ne	&$^{211}$Pb	&57.36	&27.62	&28.29	&26.44	&$1.06\times10^{-26}$&	6.85	&5.90	&44.65	&51.51	&5.86\\
$^{234}$U	&$^{26}$Ne	&$^{208}$Pb	&59.41	&25.92	&25.02	&26.61	&$1.63\times10^{-25}$&	2.20	&6.57	&50.61	&52.81	&6.60\\
$^{236}$U	&$^{26}$Ne	&$^{210}$Pb	&56.69	&$\textgreater25.9$	&30.10	&31.66	&$1.16\times10^{-27}$&	2.10	&6.42	&48.35	&50.45	&6.25\\
$^{236}$Pu	&$^{28}$Mg	&$^{208}$Pb	&79.67	&21.67	&21.48	&21.40	&$5.99\times10^{-27}$&	2.95	&7.54	&67.27	&70.22	&9.45\\
$^{238}$Pu	&$^{30}$Mg	&$^{208}$Pb	&76.80	&25.70	&25.10	&25.18	&$4.59\times10^{-28}$&	2.84	&7.33	&64.27	&67.12	&9.68\\
$^{242}$Cm	&$^{34}$Si	&$^{208}$Pb	&96.51	&23.24	&23.70	&25.46	&$3.80\times10^{-30}$&	3.57	&8.08	&79.38	&82.95	&13.56\\
\end{tabular}
\end{ruledtabular}
\end{table*}

\subsection{The Recoil Energy ($E_d$)}
A careful inspection of the $4^{th}$ column of Table 1 of Ref. \cite{josh22L} and its footnote shows that the constant parameter `$c$' (see Eq. (\Ref{P0})) is highly susceptible to shell and/or sub-shell effect and thus is higher for all the systems having $^{208}$Pb-daughter shell closure. In other words, `$c$' can be used to indicate the signature of shell and/or sub-shell closure over an isotopic chain. The shell effect is displayed in the last column of Table \Ref{tab2} such that the same cluster from different isotopes of the particular nucleus is characterized by a unique Q-value as well as the recoil energy. In such cases, the recoil energy of the daughters for double magic $^{208}Pb$ is relatively higher as compared to those of the neighbouring isotopes. Furthermore, in Fig. \Ref{fig 5} (upper panel) displays the variation of the recoil energy of the daughter nuclei w.r.t. their corresponding neutron number $N_d$ for different Q-values calculated from the RMF, $Q_{RMF}$ (dash line with blue star) and compared with the binding energies of Wang \textit{et al.} \cite{wang17} $Q_{AME}$ (solid line with black square), as well as the WS3 given by Liu \textit{et al.} \cite{liu11} $Q_{WS3}$ (dotted line with red circle). The profile shows a regular pattern for the three sets of Q-values which directly influences their magnitude. Nonetheless, in all cases, the peaks are observed for magic and/or close shell neutron number $N=126$.

Classically, one can correlate the recoil energy of the daughter nuclei ($E_d$) with the mass of the parent nuclei and the cluster in the outgoing channel. Here, we introduce a simple and intuitive systematic that governs the quantitative estimate of the recoil energy of the daughter nuclei (last column of Table \Ref{tab2}) based on the three distinct possibilities: \\
Firstly, we consider a case in which the {\bf same cluster is emitted from different parent nuclei}. From these systems, one can observe that the heavier parent produces relatively lower recoil energy provided that the daughters formed are not magic nuclei. For example, from the table, the reaction systems, namely, $^{223}$Ra $\rightarrow$ $^{14}$C+$^{209}$Pb, $^{224}$Ra $\rightarrow$ $^{14}$C+$^{210}$Pb yields $E_d$ = 2.00 and 1.91 MeV, respectively. This observation is found to be consistent provided $A_d$ is not a shell and/or sub-shell closure. Secondly, we consider {\bf emission of different clusters from the same parent nucleus}. It is observed that massive clusters produce relatively higher recoil energy. For example, the reaction systems, namely, $^{234}$U $\rightarrow$ $^{24}$Ne+$^{210}$Pb, and $^{236}$U $\rightarrow$ $^{26}$Ne+$^{210}$Pb are associated with $E_d$ = 6.03 and 6.25 MeV, respectively. Thirdly, we consider {\bf the emission of the same cluster from different parent nuclei in which at least, one of the daughters formed is a magic nucleus.} We notice that the heavier parent can produce a relatively higher or equal recoil energy. For example,  the reaction systems, namely, $^{221}$Ra $\rightarrow$ $^{14}$C+$^{207}$Pb, $^{222}$Ra $\rightarrow$ $^{14}$C+$^{208}$Pb, are yielding $E_d$ = 2.05 MeV, =2.08  MeV, respectively. A similar finding is observed for the reaction systems $^{230}$U $\rightarrow$ $^{24}$Ne+$^{206}$Pb, $^{232}$U $\rightarrow$ $^{24}$Ne+$^{208}$Pb with $E_d$ = 6.40, 6.45  MeV, respectively.

\subsection{Cluster Emission Energy ($\kappa\sqrt{Q}$)}
Besides, the dominance of the shell effect on the cluster emission energy $\kappa\sqrt{Q}$ is hinged on the formation of the double magic $^{208}$Pb daughter nucleus in which the highest peak (at $N = 126$) is formed as shown in Fig. \Ref{fig 5}(a). However, the usual peak at $N = 128$ is mainly attributed to the lower value of parameter `c' in Table \Ref{tab2}, indicating the formation of non-double magic daughter nuclei. Thus, since the shell effect is lower in such cases, the corresponding energy for the cluster tunnelling process is amply increased. On this account, we define $\kappa$ as the precise quantity/fractional amount of energy required to liberate a preformed cluster through the potential barrier. In other words, $\kappa$ is the specific amount of energy with which the preformed cluster tunnels across the Coulomb-nuclear interaction barrier. Hence, $\kappa$ can be termed as \textit{the tunnelling factor}. From Eq. (\Ref{kap}), it is evident that $\kappa$ is largely dependent on the mass of the parent, and daughter nuclei and hence, the emitted cluster.
\begin{figure*}
\includegraphics[scale=0.33]{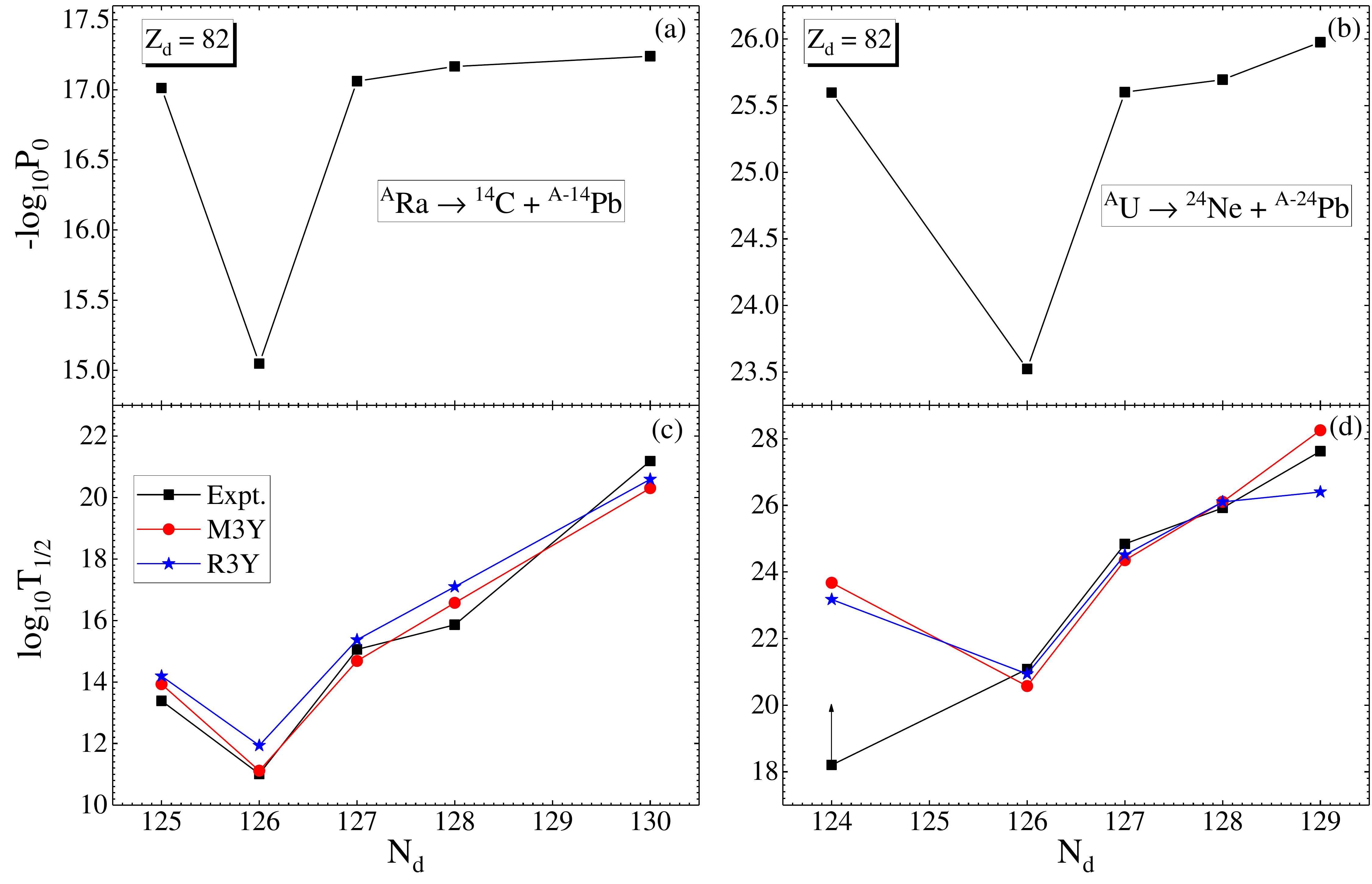}
\caption{\label{fig 6} Variation of the preformation probability $P_0$ and logarithmic half-lives $\log_{10}T_{1/2}$ with $Q_{AME}$ \cite{wang17} only for $^{A}Ra\rightarrow ^{14}C+^{A-14}Pb $ and $^{A}U\rightarrow ^{24}Ne+^{A-24}Pb$.}
\end{figure*}

\subsection{The Weighted Q-value ($bQ$)}
We reiterate that the weighted Q-value `$bQ$' is the share of the decay energy contributed during the cluster preformation process. The third panel of Fig. \Ref{fig 5} depicts the weighted Q-value of $^{24}$Ne cluster emission from even-even $^{230,232,234}$U and even-odd $^{233,235}$U isotopes as a function of the neutron number of their respective daughter nucleus. By composition, $bQ$ can only be influenced by the Q-value and parameter $b$. The predictions from the Q-values estimated from the experimental binding energies and those from the WS3 mass table relatively good agreements over the RMF deduced Q-values. For example, the difference in $Q_{AME}$ and $Q_{WS3}$ ranges between 0.1 - 0.5 MeV, whereas the $Q_{RMF}$ with a difference of about 1.0 - 3.0 MeV. However, despite the variation of the Q-values, the behaviour of the even(Z)-even(N) systems are different from those of the even(Z)-odd(N) systems. In other words,  Fig. \Ref{fig 5}c shows that all the even-even systems have lower $bQ$ values and follow the same trend, unlike the even-odd systems which are marked with higher values.  As such, it is apparent that parameter $b$ captures the odd-even staggering effect since the presence of unpaired protons or neutrons in the open-shell radioactive cluster emitters are uniquely associated with higher $b$ values (at $N=127$ and $N=129$) and thus influences the preformation probability. This fully agrees with the systematic study in Ref. \cite{seif15}. Details of the pairing effect for open-shell nuclei, the odd-even staggering effects as well as their connection with the single-particle energies and orbital filling have been extensively discussed in Ref. \cite{bhu21,an22,good21,kosz21}. Therefore, it is evident that a careful determination of $P_0$ provides ample information about the nuclear structure and the kinematics of cluster emission. Using Eq. (\Ref{P0}) a detailed calculation of the cluster preformation is carried out for heavy nuclei decaying to the double-shell closure $^{208}$Pb-daughters and its neighbours. Thus, the credibility of the estimated $P_0$ is graphically illustrated and discussed. Since the preformation probability is not a direct experimentally observed quantity, one relies on the theoretical models for its deduction.

\subsection{Cluster Preformation $P_0$ and Half-lives $T_{1/2}$}
Figure \Ref{fig 6} (upper panels (a) and (b)) displays the calculated preformation probability of $^{14}$C and $^{24}$Ne clusters from Ra and U isotopic chains respectively as a function of the neutron number of their corresponding  Pb-daughter nuclei formed. From the figure, it is apparent that cluster preformation in heavy nuclei is usually accompanied by the appearance of a notable dip at $N=126$. This  agrees with the statistical analysis of Bonetti \textit{et al.} \cite{bone74} and Hodgson \textit{et al.} \cite{hodg03} in $\alpha$-particle preformation. The figure further asserts that the $P_0$ cannot maintain a constant magnitude for different nuclei, unlike the conjecture in Refs. \cite{xu06,qian11}. Thus, in both figures, the $-\log P_0$ values for similar cluster emissions from different isotopes of an element have a clear distinction. Taking the inverse of  $-\log P_0$  values in  Fig. \Ref{fig 6}(a) and Fig. \Ref{fig 6}(b), the $P_0$ fall in the range $0<P_0<1$. By implication, the treatment of $P_0$ as unity relegates the exclusive properties of the participating nuclei. A detailed inspection of both figures (and  column 8 of Table \Ref{tab2}) for $^{A}Ra\rightarrow ^{14}C+^{A-14}Pb$  and $^{A}U\rightarrow ^{24}Ne+^{A-24}Pb $ reveals that the cluster preformation probability $P_0$ at the double magic shell closure $^{208}$Pb daughter is higher than those its neighbouring daughter nuclei with about an order of $10^2$ and exhibits relatively lower half-lives. This reflects the stability of  $^{222}$Ra and $^{232}$U parents against $^{14}$C and $^{24}$Ne cluster decays respectively.  

Similarly, the theoretically determined half-lives using the M3Y and R3Y potentials are compared with the experimental data .
As representative cases, Fig. \Ref{fig 6}(c) and Fig. \Ref{fig 6}(d) shows the variation of the logarithmic half-lives $\log_{10} T_{1/2}$  for $^{14}$C and $^{24}$Ne cluster decay from Ra and U isotopes respectively as a function of the neutron number of the daughter ($N_d$). In both instances, the minima in $\log_{10} T_{1/2}$ are identified with the decay leading to the double magic daughter $^{208}$Pb ($Z_d=82$ and $N_d=126$). In other words, the shell stabilises at magic daughter nuclei. This infers that the stability of the cluster emitters could be explained via the shell closure effect. Besides, the half-lives of cluster emitters appear to increase w.r.t. the magnitude of the neutron number of the daughter nuclei formed until a magic number (or its neighbour) is attained. The M3Y and R3Y predictions are found to be consistent with the experimentally observed half-lives for all the systems under study. However, the cluster-decay half-lives of certain systems (like for $N_d=124$ in Fig. \Ref{fig 6}(d)) lack precise experimental measurement, for which only the lower limits were given (marked with an upward arrow). In that vein, the prediction of M3Y and R3Y agrees with the experimental lower limit and can be considered to be more probable from the theoretical point of view (since the deepest minima are usually associated with the double magic number $N_d=126$). The slight difference in the estimation of the M3Y and R3Y NN potentials reflects the uniqueness of their barrier characteristics.

\section{SUMMARY AND CONCLUSIONS}
\label{summary}
The proposed new preformation probability $P_0$ formula captures several known theoretically established factors (such as the cluster mass $A_c$, mass and charge asymmetry ($\eta_A$ and $\eta_Z$) and the Q-value) affecting the mechanism and kinematics of the cluster emissions is applied here. Also, for the first time, an expression for the contribution of the decay energy in terms of the cluster preformation, emission and recoil energy is applied in the cluster radioactivity. Further, we present a new set of criteria for estimating the recoil energy of daughter nuclei in the cluster radioactivity. Using the relativistic mean-field (RMF) approach, the cluster decay half-lives of various nuclei decaying to (or around) $^{208}$Pb-daughters are calculated within the preformed cluster-decay model (PCM). As a result, the stability of the cluster emitters is closely linked with the pairing and shell closure effect. Although the barrier properties of the phenomenological M3Y and microscopic R3Y NN potentials differ qualitatively, their respective predictions are found to relatively agree with the experimentally measured half-lives. To gain a comprehensive insight into the cluster decay dynamics, it is of future interest to systematically investigate the preformation properties and half-lives of systems yielding non-Pb daughters along with various proximity potentials with deformation and orientation effects. This study can also be extended to predict cluster radioactivity in the unknown territories of the superheavy mass region.

\section*{acknowledgments}
The authors would like to acknowledge the support from the Fundamental Research Grant Scheme (FRGS) under the grant number FRGS/1/2019/STG02/UNIMAP/02/2 from the Ministry of Education Malaysia stipulated with the Institute of Engineering Mathematics (IMK), UniMAP as the beholder, Science and Engineering Research Board (SERB), File No. CRG/2021/001229, FOSTECT Project Code: FOSTECT.2019B.04, and FAPESP Project Nos. 2017/05660-0.\\


\begin{thebibliography}{99}
\bibitem{sand80}
A.  Sandulescu,  D. N.  Poenaru,    and  W.  Greiner,  Sov.  J. Part. Nucl.(Engl. Transl.);(United States) {\bf 11} (1980).
\bibitem{gupt94}
R. K. Gupta and W. Greiner, Int. J. Mod. Phys. E \textbf{3}, 335 (1994, Suppl.).
\bibitem{rose84}
 H. J. Rose and G. A. Jones, Nature {\bf 307}, 245 (1984).
\bibitem{bone07}
R. Bonetti and A. Guglielmetti, Rom. Rep. Phys. \textbf{59}, 301
(2007).
\bibitem{qian12}
Y. Qian and Z. Ren, J. Phys. G \textbf{39}, 015103 (2012).
\bibitem{deng15}
D. Deng, Z. Ren, D. Ni, and Y. Qian, J. Phys. G \textbf{42}, 075106 (2015).
\bibitem{hoos05}
M. A. Hooshyar, I. Reichstein, and F. B. Malik, Nuclear Fission and Cluster Radioactivity: An Energy-Density Functional Approach (Springer-Verlag, Berlin, 2005).
\bibitem{poen85}
D. N. Poenaru, M. Ivascu, A. Sandulescu, and W. Greiner, Phys. Rev. C \textbf{32}, 572 (1985).
\bibitem{poen86}
D. N. Poenaru, W. Greiner, K. Depta, M. Ivascu, D. Mazilu, and A. Sandulescu, At. Data Nucl. Data Tables \textbf{34}, 423 (1986).
\bibitem{mali89}
S. S. Malik and R. K. Gupta, Phys. Rev. C \textbf{39}, 1992 (1989).
\bibitem{gupt88}
R. K. Gupta, in \textit{Proceedings of the 5th International Conference on Nuclear Reaction Mechanisms}, edited by E. Gadioli (Ricerca Scientifica ed Educazione Permanente, Milan, 1988), p. 416.
\bibitem{wei17}
K. Wei and H. F. Zhang, Phys. Rev. C \textbf{96}, 021601(R) (2017).
\bibitem{blen88}
R. Blendowske and H. Walliser, Phys. Rev. Lett. \textbf{61}, 1930 (1988).

\bibitem{sant21c}
K. P. Santhosh and Tinu Ann Jose, Phys. Rev. C \textbf{104}, 064604 (2021).
\bibitem{ni10}
D. Ni and Z. Ren, Phys. Rev. C \textbf{82}, 024311 (2010).
\bibitem{bala14}
M. Balasubramaniam and N. S. Rajeswari, Int. J. Mod. Phys. E \textbf{23}, 1450018 (2014).
\bibitem{deng14}
D. Deng  and Z. Ren, Phys. Rev. C \textbf{93}, 044326 (2016).
\bibitem{josh22L}
T. M. Joshua,  R. Kumar, and M. Bhuyan, Submitted to Physical Review Letters (2022).
\bibitem{ropk14}
G. R{\"o}pke, P. Schuck, Y. Funaki, H. Horiuchi, Zhongzhou Ren, A. Tohsaki, Chang Xu, T. Yamada, and Bo Zhou, Phys.
Rev. C \textbf{90}, 034304 (2014).
\bibitem{xu16}
C. Xu, Z. Ren, G. R{\"o}pke, P. Schuck, Y. Funaki, H. Horiuchi, A. Tohsaki, T. Yamada, and B. Zhou, Phys. Rev. C \textbf{93}, 011306 (2016).
\bibitem{levi53}
J. S. Levinger, Phys. Rev. \textbf{90}, 11 (1953).
\bibitem{stra01}
G. Stratan, W. Scheid, Int. J. Mod. Phys. E \textbf{10}, 367 (2001).
\bibitem{kuma12c}
R. Kumar, Phys. Rev. C \textbf{86}, 044612 (2012).
\bibitem{quen78}
P. Quentin and  H. Flocard, Annu. Rev. Nucl. Part. Sci. {\bf 28}, 523 (1978).
\bibitem{horn75}
W. H. Hornyak, Nuclear Structure (Academic Press, New York,
1975).
\bibitem{schu16}
N.  Schunck  and  L.  Robledo,  Rep. Prog. Phys. {\bf 79}, 116301 (2016).
\bibitem{vaut72}
D. Vautherin and D. M. Brink, Phys. Rev. C {\bf 5}, 626 (1972).
\bibitem{epel09}
E. Epelbaum, H.-W. Hammer, and U.-G. Mei${\ss}$ner, Rev. Mod. Phys. {\bf 81}, 1773 (2009).
\bibitem{ekst13}
A. Ekstr$\ddot{o}$m {\it et al.}, Phys. Rev. Lett. {\bf 110}, 192502 (2013).
\bibitem{sing12}
B. B. Singh, M. Bhuyan, S. K. Patra,  and R. K. Gupta, J. Phys. G: Nucl. Part. Phys. {\bf 39}, 069501 (2012).
\bibitem{sing10}
B. B. Singh, M. Bhuyan, S. K. Patra,  and R. K. Gupta, arXivpreprint arXiv:1011.5732  (2010).
\bibitem{satc79}
G. R. Satchler and W. G. Love, Phys. Rep. {\bf 55}, 183 (1979).
\bibitem{Bisw20}
S. Biswal, M. A. El Sheikh, N. Biswal, N. Yusof, H. Kas-sim, S. K. Patra,  and M. Bhuyan, Nucl. Phys. A {\bf 1004}, 122042 (2020).
\bibitem{Itag20}
N. Itagaki, A. Afanasjev, and D. Ray, Phys. Rev. C {\bf 101}, 034304 (2020).
\bibitem{Tani20}
A. Taninah, S. Agbemava, and A. Afanasjev, Bull. Am. Phys. Soc. \textbf{65} (2020).
\bibitem{liu11}
Min Liu, Ning Wang, Yangge Deng, Xizhen Wu, Phys. Rev. C {\bf 84}, 014333 (2011).
\bibitem{wang17}
M. Wang, G. Audi, F. G. Kondev, W. J. Huang, S. Naimi,
and X. Xu, Chin. Phys. C \textbf{41}, 030003 (2017).

\bibitem{jos22a}
T. M. Joshua, N. Jain, R. Kumar, K. Anwar, N. Abdullah, and M. Bhuyan, Foundations, \textbf{2}, 85-104 (2022).
\bibitem{ring96}
P. Ring, Prog. Part. Nucl. Phys. \textbf{37}, 193 (1996).
\bibitem{bhu15}
M. Bhuyan, Phys. Rev. C \textbf{92}, 034323 (2015).
\bibitem{bhu20}
M. Bhuyan, R. Kumar, S. Rana, D. Jain, S. K. Patra, and
B. V. Carlson, Phys. Rev. C \textbf{101}, 044603 (2020).
\bibitem{sero86}
B. D. Serot and J. D. Walecka, Adv. Nucl. Phys. \textbf{16}, 1 (1986).
\bibitem{sing22}
A. Singh, A. Shukla,  and M. K.  Gaidarov, J. Phys. G: Nucl. Part. Phys, \textbf{49}, 025101 (2021).

\bibitem{sing11}
B. B. Singh, S. K. Patra, and R. K. Gupta, Phys. Rev. C 82,
014607 (2010); Int. J. Mod. Phys. E \textbf{20}, 1003 (2011).

\bibitem{bone99}
R. Bonetti and A. Guglielmetti, in Heavy elements and related new phenomena
edited by R. K. Gupta and W. Greiner (World Scientific Pub., Singapore, 1999)
Vol. 2, p. 643.

\bibitem{deli09}
D. S. Delion, Phys. Rev. C \textbf{80}, 024310  (2009). 

\bibitem{isma14}
M. Ismail, A. Adel, Phys. Rev. C \textbf{89}, 034617 (2014).
\bibitem{bhuy18}
M. Bhuyan and R. Kumar, Phys. Rev. C \textbf{98}, 054610
(2018).
\bibitem{grei86}
M. Greiner and W. Scheid, J. Phys. G: Nucl. Phys. \textbf{12}, L229
(1986).
\bibitem{shar19}
K. Sharma, G. Sawhney, M. K. Sharma, R. K. Gupta, Eur. Phys. J. A \textbf{55}, 30 (2019).
\bibitem{bloc77}
J. Blocki, J. Randrup, W. J. Swiatecki, and C. F. Tsang, Ann. Phys. (N.Y.) \textbf{105}, 427 (1977).
\bibitem{niyt15}
Niyti, G. Sawhney, M. K. Sharma, and R. K. Gupta, Phys. Rev. C \textbf{91}, 054606 (2015).
\bibitem{shar18}
K. Sharma, G. Sawhney, M. K. Sharma, and R. K. Gupta, Nucl. Phys. A \textbf{972}, 1 (2018).
\bibitem{jos21c}
T. M. Joshua, K. Anwar, N. Abdullah, N. Jain, S. Rana, R. Kumar, M. Bhuyan, Proceedings of the DAE-BRNS symposium on nuclear physics. V. \textbf{65} (2021).
\bibitem{jos22b}
J. T. Majekodunmi,  M. Bhuyan, D. Jain, K. Anwar, N. Abdullah, and R. Kumar, Phys. Rev. C  \textbf{105}, 044617 (2022).
\bibitem{bhu21}
M. Bhuyan, B. Maheshwari, H. A. Kassim, N. Yusof, S. K. Patra, B. V. Carlson and P. D. Stevenson, J. Phys. G: Nucl. Part. Phys. \textbf{48}, 075105 (2021).
\bibitem{an22}
R. An, X. Jiang, L.-G. Cao, and F.-S. Zhang, Phys. Rev. C \textbf{105}, 014325 (2022).
\bibitem{good21}
T. D. Goodacre \textit{et al.}, Phys. Rev. C \textbf{104}, 054322 (2021).
\bibitem{kosz21}
Koszor{\'u}s, {\'A} \textit{et al.}, Nat. Phys. \textbf{17}, 439 (2021).
\bibitem{seif15}
W. Seif, Phys. Rev. C \textbf{91}, 014322 (2015).
\bibitem{bone74}
R. Bonetti and L. Milazzo-Colli, Phys. Lett. B \textbf{49}, 17 (1974).
\bibitem{hodg03}
P. E. Hodgson and E. B{\v{e}}t{\'a}k, Phys. Rep. \textbf{374}, 1 (2003).
\bibitem{xu06}
C. Xu and Z. Ren, Phys. Rev. C \textbf{74}, 014304 (2006); \textbf{73},
041301(R) (2006).
\bibitem{qian11}
Y. Qian, Z. Ren, and D. Ni, Phys. Rev. C \textbf{83}, 044317 (2011).

\end{thebibliography}
\end{document}